\documentstyle[preprint,tighten,floats,eqsecnum,aps,epsfig]{revtex}

\begin{document}
\title{ 
A statistical theory of the mean field
}
\author{ 
 R. Caracciolo$^a$, A. De Pace$^a$, H. Feshbach$^b$ and A. Molinari$^a$
}
\address{
 ${}^{a}$ Dipartimento di Fisica Teorica dell'Universit\`a di Torino and \\
Istituto Nazionale di Fisica Nucleare, Sezione di Torino, \\
 via P.Giuria 1, I-10125 Torino, Italy \\
 $^b$ Center for Theoretical Physics, \\
 Laboratory for Nuclear Science and Department of Physics, \\
 Massachusetts Institute of Technology, Cambridge, MA 02139, USA
}

\maketitle

\begin{abstract}
A statistical theory of the mean field is developed.
It is based on the proposition that the mean field can be obtained as an energy
average. Moreover, it is assumed that the matrix elements of the residual
interaction, obtained after the average interaction is removed, are random 
with the average value of zero.
With these two assumptions one obtains explicit expressions for the mean field
and the fluctuation away from  the average.
The fluctuation is expanded in terms of more and more complex excitations.
Using the randomness of the matrix elements one can then obtain formulas 
for the contribution to the error from each class of complex excitations and a
general condition for the convergence of the expansion is derived.
It is to be emphasized that no conditions on the nature of the system being
studied are made.
Making some simplifying assumptions a schematic model is developed.
This model is applied to the problem of nuclear matter. The model yields a measure
of the strength of the effective interaction.
It turns out to be three orders of magnitude less than that calculated using a
potential which gives a binding energy of about $-7 \, {\rm MeV}/{\rm nucleon}$
demonstrating the strong damping of the interaction strength induced by the
averaging process.
\end{abstract}

\vfill
\begin{center}
Submitted to: {\em Annals of Physics (N.Y.)}
\end{center}

\vfill

\noindent
DFTT-30/97 \hfill \\
nucl-th/9705044 \hfill May 1997

\eject
                             
\section{Introduction}
\label{sec:intro}
Remarkably, the mean field provides a good description of many systems. The
underlying dynamics may be very different, the interparticle interaction may be
long range or short range; it may be weak or strong.
Nevertheless in each case a mean field description agrees closely with
experiment. The question naturally arises: what are the common features which
these various systems have which ensure the effectiveness of the mean field
approximation?
This is the issue which mostly concerns us in this paper.

We proceed by presenting a formalism in which the mean field and the error are
derived without any reference to a particular dynamical system. The mean field
is obtained as an energy average which smooths over short time events. This
procedure also develops an expression for the error which must be small if the
mean field is to be useful. To calculate the error  a related assumption is
used. Not only is an energy average taken, but it is also assumed that the error
matrix elements are random with the consequence that the average error is zero,
but of course the average of the square of the error is not.
The error can be calculated as a finite series consisting of contributions from
various degrees of excitation. A general expression for each term is developed
and a parameter determining the relative importance of each term is identified.

It will be noted that the assumptions listed above, namely energy averaging and
random error matrix elements, are identical with those used in the statistical
theory of nuclear reactions \cite{Fes92}.

This paper concludes  with an exploratory calculation appropriate for the
ground state of nuclear matter. Only the first order error is considered.
One obtains reasonable estimates of the error. But importantly there is an
enormous reduction in the strength of the residual interaction from that
obtained from a Fermi gas model which must be a consequence of the statistical
approach.

The statistical approach provides a method  for calculating the slowly varying
and rapidly varying contributions to physical observables. It differs from the
customary approach based on the Pauli principle and dynamical correlations. The
latter are a consequence of the rapidly varying contributions. The theory
presented in this paper is essentially a different way of taking these
contributions into account.

Our work is organized as follows. In sect.~\ref{sec:formalism} the formalism is 
revisited, --- in particular the partition of the Hilbert space into the $P$ and
$Q$ sectors, --- and the related equations are discussed, while in 
{sect.}~\ref{sec:energy-average} a precise definition of the {\em energy 
average} is provided. The latter is performed with a 
distribution, characterized by a single parameter $\epsilon$, amounting to shift
the position of the poles corresponding to the excited states of the system away
from the ground state energy.
In the same section, we pave the way for setting up a suitable formalism to
account for the corrections (the error) to the mean field energy.

In sect.~\ref{sec:corrections} the expression for the corrections to the mean
field energy $\bar{E}_0$ is given in the framework of the expansion, above referred to, 
built out of successive 2p--2h excitations. 

In sect.~\ref{sec:HF} we deal with the issue of introducing an operator,
projecting into a smooth $P$-space, simple enough to allow the present approach 
to be worked out for nuclear matter. We thus restrict the $P$-space, --- also 
with the aim of rendering transparent the comparison between the mean field 
energy of the Hartree-Fock (HF) and of the present theory, --- just to a simple state: The HF
determinant. With this choice for $P$, we are able to establish an expression
connecting the true ground state $E$, the mean field $\bar{E}_0$ and the Hartree-Fock 
$E_{\text{HF}}$ energies. Importantly, $\bar{E}_0$ is shown to be {\em always}
lower than $E_{\text{HF}}$ and it turns out to be proportional to the variance,
to be later defined, of the residual effective interaction $V$ among nucleons,
which, in our context, is precisely defined.

In sect.~\ref{sec:spectr-fact} we address the issue of the
occupation probability $S^2$ (spectroscopic factor) of the ground state of
nuclear matter as defined by the structure of our $P$-space.
We are able to deduce, within our scheme, an expression for $S^2$ in terms of
$E$, $\bar{E}_0$ and $E_{\text{HF}}$, and moreover embodying the energy
derivative of $V$.

Finally, in sect.~\ref{sec:numerical-results} 
a scheme which permits numerical predictions of the
present approach is developed. 

 For sake both of simplicity and of illustration, we
account for the first corrections only and, even so, we are
forced to introduce a free parameter $\alpha$ in order to give an estimate of
their size.
We are then able to set up two systems of two equations each by
coupling the equation for the mean field energy to the one for the fluctuations.
By solving the systems and with a convenient choice of the two parameters
entering into our approach, namely $\epsilon$ and $\alpha$, 
we obtain results consistent with the empirical features of nuclear matter
as obtained through an extrapolation from finite nuclei.
We provide at the same time an estimate for the fluctuations of the mean
field energy.
These goals are achieved with a residual effective interaction which, 
as previously emphasized, is lowered by about three orders of magnitude with respect to the bare one
and yields a spectroscopic factor not in conflict with other independent estimates \cite{Fant}.

In the concluding section our results are summarized and possible improvements discussed.

\section{Formalism}
\label{sec:formalism}

In this section we shortly revisit the formalism introduced in 
Ref.~\cite{Fes96}.
Let $H$ be the nuclear hamiltonian entering into the Schr\"odinger equation
\begin{equation}
 H \Psi = E \Psi\,  . 
\label{mascheq}
\end{equation}
Let $P$ be the hermitian operator projecting the nuclear ground state into the 
Hilbert subspace of functions associated with the low momentum transfer physics
and $Q$ the operator complementing $P$. Clearly
\begin{equation}
P^2 = P,\qquad Q^2 = Q,\qquad PQ = 0,\qquad P + Q = {\bf 1} \, . 
\label{project}
\end{equation}
It is then proved that the pair of equations
\begin{equation}
(E - H_{PP} )\, (P\Psi )= H_{PQ}\, (Q\Psi )\,  , 
\label{sist1}
\end{equation}
and 
\begin{equation}
(E - H_{QQ} )\, (Q\Psi )= H_{QP}\, (P\Psi )\,  ,
\label{sist2}
\end{equation}
are equivalent to (\ref{mascheq}). In the above 
formulas the following shorthand
notations have been introduced, viz.
\begin{equation}
H_{PP} = PHP,\qquad H_{QQ} = QHQ,\qquad H_{PQ} = PHQ,\qquad
H_{QP} = QHP\,  .
\label{notaz}
\end{equation}
Let us  next define an auxiliary function $\Psi_0$, 
which in the end will disappear from
the formalism, such that
\begin{equation}
(E - H_{PP} )\, \Psi_0 = 0 \, . 
\label{eqpsi0}
\end{equation}
Then, from (\ref{sist1}) it follows
\begin{equation}
(P\Psi ) = \Psi_0 + \frac{1}{E - H_{PP}}\,  H_{PQ}\, (Q\Psi ) \, , 
\label{ppsieq}
\end{equation}
which, when inserted into (\ref{sist2}), yields
\begin{equation}
(Q\Psi ) =\frac{1}{E - H_{QQ} - W_{QQ}}\, H_{QP} \Psi_0 \equiv
 \frac{1}{e_Q}\,  H_{QP}\, \Psi_0 \, ,
\label{qpsieq}
\end{equation}
being
\begin{equation}
W_{QQ} = H_{QP}\, \frac{1}{E - H_{PP}}\, H_{PQ} 
\label{wqqdef}
\end{equation}
and, obviously, 
\begin{equation}
e_{Q} = E - H_{QQ} - W_{QQ} \, .
\label{eqdef}
\end{equation}
By using (\ref{ppsieq}) and (\ref{qpsieq}) one can, on the one hand, recast 
(\ref{sist1}) as follows
\begin{equation}
(E - H_{PP} )\, (P\Psi )= H_{PQ}\,  \frac{1}{e_Q}\,  H_{QP}\, \Psi_0 \, ,
\label{sist1psi0}
\end{equation}
and, on the other, express $\Psi_0$ according to
\begin{equation}
\Psi_0 = \frac{1}{1 + 
\frac{\strut\displaystyle 1}{\strut\displaystyle E - H_{PP}}\, H_{PQ}\, 
\frac{\strut\displaystyle 1}{\strut\displaystyle e_Q}\, H_{QP}}
\, (P\Psi ) \, .
\label{psi0eq}
\end{equation}
The insertion of (\ref{psi0eq}) into (\ref{sist1psi0}) 
leads in turn to the
equation
\begin{equation}
(E - H_{PP} )\, (P\Psi )= H_{PQ}\,  \frac{1}{e_Q}\,  H_{QP}\, 
\frac{1}{1 + \frac{\strut\displaystyle 1}{\strut\displaystyle E - H_{PP}}\, 
  H_{PQ}\, \frac{\strut\displaystyle 1}{\strut\displaystyle  e_Q}\, 
H_{QP}}\  (P\Psi ) \, ,
\label{sist1exp}
\end{equation}
which can be reorganized into the form
\begin{eqnarray}
(E - H_{PP} )\, (P\Psi ) & = & 
H_{PQ}\,  \frac{1}{e_Q}\, 
\frac{1}{1 + H_{QP}\, 
\frac{\strut\displaystyle 1}{\strut\displaystyle E - H_{PP}}\, H_{PQ}\, 
\frac{\strut\displaystyle 1}{\strut\displaystyle e_Q} }\,
H_{QP}\, (P\Psi )  \nonumber \\
 & = &  H_{PQ}\,  \frac{1}{e_Q}\, 
\frac{1}{1 + W_{QQ}\, 
\frac{\strut\displaystyle 1}{\strut\displaystyle e_Q}}\, H_{QP}\, (P\Psi )  \, .
\label{sist1exp1}
\end{eqnarray}
In deriving (\ref{sist1exp1}) the operator identity
\begin{equation}
B\, \frac{1}{1 + CAB} = \frac{1}{1 + BCA}\, B 
\label{opident}
\end{equation}
has been used, setting
\begin{equation}
A = \frac{1}{e_Q}\,\  , \qquad 
B = H_{QP} \, ,\qquad C =  \frac{1}{E - H_{PP}}\,  H_{PQ} \, .
\label{abcdef}
\end{equation}
Then, since
\begin{equation}
 \frac{1}{e_Q}\, \frac{1}{1 + W_{QQ}\, 
\frac{\strut\displaystyle 1}{\strut\displaystyle e_Q}}\,  = 
\frac{1}{\left (
\frac{\strut\displaystyle 1}{\strut\displaystyle e_Q}\right )^{-1}\,   
+ W_{QQ} }\, , 
\label{1eqid}
\end{equation}
one finally obtains the equation
\begin{equation}
\left (E - H_{PP} - H_{PQ}\, 
\frac{1}{\left (
\frac{\strut\displaystyle 1}{\strut\displaystyle e_Q}\right )^{-1}\,  
+ W_{QQ}}\, H_{QP}\right )\,
(P\Psi ) = 0 \, .
\label{fesheqa}
\end{equation}

We thus see that the intricate many-body operator
\begin{equation}
{\cal H} = H_{PP}  + H_{PQ}\,
\frac{1}{\left (
\frac{\strut\displaystyle 1}{\strut\displaystyle e_Q}\right )^{-1}\,   
+ W_{QQ}}\, H_{QP}\, ,
\label{calh}
\end{equation}
dependent upon the {\em exact} ground state energy of the system $E$, defines an
equation which is not an eigenvalue equation in the usual sense. 
Indeed, the dependence upon $E$ is non-linear, since the latter appears in the 
propagator in the $Q$-space in (\ref{fesheqa}).

\section{Energy average}
\label{sec:energy-average}

As emphasized in the introduction, $(P\Psi )$ contains both slow and rapidly
varying components. To obtain the slowly varying components  one must perform an
energy average of the wave function. The components varying most rapidly with
energy are taken to be in the $Q$ space so that $(Q\Psi )$ is to be replaced by 
$\langle  Q\Psi \rangle\ $.

As a consequence of this procedure also the $(P\Psi )$ component of the wave
function and the energy $E$ as well, will no longer be the same 
(see eq.~(\ref{sist1})). We shall call the modified quantities 
$\langle  P\Psi \rangle\ $ and $\bar E_0$ respectively and obtain, instead of 
(\ref{ppsieq}), the following equation
\begin{equation}
\langle  P\Psi \rangle\  = \tilde \Psi_0 + \frac{1}{{\bar E_0} 
- H_{PP}}\,
  H_{PQ}\, \langle Q\Psi \rangle\ \,  , 
\label{avppsieq}
\end{equation}
where
\begin{equation}
\left ( {\bar E_0} - H_{PP} \right ) \tilde \Psi_0 = 0 \, .
\label{phi0eq}
\end{equation}
Analogously, eq.~(\ref{qpsieq}) will now read 
\begin{equation}
\langle Q\Psi \rangle\ = \langle 
\frac{1}{e_Q}\rangle\ \, H_{QP} \tilde \Psi_0 \, ,
\label{avqpsieq}
\end{equation}
which, upon insertion into (\ref{avppsieq}), leads to
\begin{equation}
\left ( {\bar E_0} - H_{PP} \right ) \langle P\Psi \rangle\  =
H_{PQ}\, \langle \frac{1}{e_Q}\rangle\ \, H_{QP} \tilde \Psi_0  \, .
\label{avppsiphi0}
\end{equation}
Exactly as before, $\tilde \Psi_0$  can now be obtained 
by combining (\ref{avppsieq}) with
 (\ref{avqpsieq})  and, when inserted into
 (\ref{avppsiphi0}),  it yields
\begin{equation}
\left ( {\bar E_0} - H_{PP} 
- H_{PQ} \frac{1}{{\langle 
\frac{\strut\displaystyle 1}{\strut\displaystyle e_Q} \rangle}^{-1} \, 
+ W_{QQ}} \, H_{QP}
\right )\,  \langle P\Psi \rangle\  \equiv 
\left ( {\bar E_0} - {\bar {\cal H}}\right )\, 
\langle P\Psi \rangle\  = 0  \, ,
\label{avfesheq}
\end{equation}
which is the correspondent of equation (\ref{fesheqa}) when a suitable energy 
average has been performed.

At this point we shall specify how the energy average is to be carried out. One
introduces a smoothing function $\rho (E , {\bar E_0} )$ with the property
\begin{equation}
\int \rho (E , {\bar E_0} ) \, dE = 1 \, .  
\label{intnew}
\end{equation}
The average of a function $f(E)$ is then 
\begin{equation}
\langle f \rangle  = \int \rho (E , {\bar E_0} ) \, f(E) dE \, .
\label{avef}
\end{equation}
Since we are considering bound states we {\em require}
 the average $\langle f \rangle$
to be real. A smoothing function which satisfies these conditions is 
\begin{equation}
  \rho (E , \bar E_0 ) = \frac{1}{2\pi \text{i}} 
    \, \frac{1}{E-(\bar E_0-\epsilon)} \, 
\label{rhoreal}
\end{equation}
with a path of integration in eq.~(\ref{avef}) going along the real axis $Re E$ with a small
semi--circle described positively about the singularity 
$({\bar E_0} - \epsilon)$. Assuming boundedness  conditions at infinity
sufficiently strong, the Cauchy's integral formula can be applied so that the
condition of eq.~(\ref{intnew}) is satisfied and eq.~(\ref{avef}) becomes 
\begin{equation}
\langle f \rangle  = f ({\bar E_0} -\epsilon) \, .
\label{avefeps}
\end{equation}

 Following the procedure developed by Kawai, Kerman and McVoy \cite{kaw75}, we find 
\begin{eqnarray}
(E - \bar {\cal H} )\, (P\Psi ) & = &  V_{PQ}\, (Q\Psi )\,  
\label{avsista} \\
(E - H_{QQ} )\, (Q\Psi ) & = &  V_{QP}\, (P\Psi )\,  
\label{avsistb}
\end{eqnarray}
where 
\begin{equation}
\bar {\cal H} = H_{PP} + V_{PQ} V_{QP} \, 
\frac{1}{\bar E_0  -\epsilon - E} \, 
\label{hbarv}
\end{equation}
and  
\begin{eqnarray}
V_{PQ} & = &  H_{PQ}\ \sqrt{\frac{\bar E_0 - \epsilon -E}
{\bar E_0  - \epsilon - H_{QQ}} } \, \, , 
\label{vpqdefr} \\
V_{QP} & = &  \sqrt{\frac{\bar E_0  - \epsilon -E}
{\bar E_0  - \epsilon - H_{QQ}} } \,\,  H_{QP} \, .
\label{vqpdefr}
\end{eqnarray}
Note that eqs.~(\ref{avsista}) and (\ref{avsistb}) 
are of the same form as the original 
eqs.~(\ref{sist1}) and (\ref{sist2}). However the potential which couples 
$(P\Psi )$ and $(Q\Psi )$ is $V_{PQ}$ rather than $H_{PQ}$. The strength of the
coupling is thus considerably reduced by roughly 
$(\epsilon / H_{QQ})^{\frac{1}{2}}$ which is much less than one. This assumes
that $H_{PP}$ and $\bar {\cal H}$ are of the same order of magnitude.

Concerning the parameter $\epsilon$, in principle its value should be inferred from
experiment; in practice we shall consider it as a free parameter to be utilized
in the search for agreement with experiment. 

Now  the spectral decomposition of the operator $1/( E - \bar {\cal H})$ can 
be performed:
\begin{eqnarray}
(P\Psi ) & = & \frac{1}{E - \bar {\cal H}}\, V_{PQ} (Q\Psi ) = 
\sum_{n} \frac{|\ \Phi_n\ \rangle \langle\  \Phi_n\ |}{E - \bar E_n} \,
V_{PQ} (Q\Psi ) \nonumber\\
& = & |\ \Phi_0\ \rangle\, 
\frac{\langle\ \Phi_0\ |\ V_{PQ}\  |\ Q\Psi\  \rangle}{E - \bar E_0} +
\left ( \frac {1}{E - \bar {\cal H}}\right )^\prime\, 
V_{PQ} (Q\Psi ) \, ,
\label{spectd}
\end{eqnarray} 
where
\begin{equation}
\bar {\cal H} \Phi_n = \bar E_n \Phi_n \, .
\label{hbareig}
\end{equation}
The prime on $(1/( E - \bar {\cal H}))$
signify that the lowest eigenfunction $\Phi_0$ is to be excluded. 

 From eq.~(\ref{spectd}) we have 
\begin{equation}
\langle\ \Phi_0\ |\  P\Psi \  \rangle = 
\frac{\langle\ \Phi_0\ |\ V_{PQ}\  |\ Q\Psi\  
\rangle}{E - \bar E_0}  \, .
\label{normc}
\end{equation}
To obtain $(Q\Psi )$ we return to eqs.~(\ref{avsista}) and (\ref{avsistb}) and get
\begin{equation}
(Q\Psi ) = \frac{1}{E - h_{QQ}}\, V_{QP}\, |\Phi_0\rangle 
\langle \Phi_0 | P\Psi \rangle \, ,
\label{qpsiphi0}
\end{equation}
where 
\begin{equation}
h_{QQ} = H_{QQ} + \bar W_{QQ} \, 
\label{shqq}
\end{equation}
and 
\begin{equation}
\bar W_{QQ} \equiv V_{QP}\, \left ( \frac {1}{E - 
\bar {\cal H}}\right )^\prime\,
V_{PQ} \, .
\label{wbardef}
\end{equation}
Inserting eq.~(\ref{qpsiphi0}) into eq.~(\ref{normc}) one obtains
\begin{equation}
 E - \bar E_0  = \langle\ \Phi_0\ |\ V_{PQ}\,
\frac{1}{E - h_{QQ}}\, V_{QP}\ |\ \Phi_0\ \rangle  \, .
\label{deltae}
\end{equation}
In this equation $\bar E_0$ is the mean field energy while the right hand side
represents the correction.

Note that the quantity  $\langle \Phi_0 | P\Psi \rangle$ does not enter into 
eq.~(\ref{deltae}). This is to be expected since, so far, the amplitude of 
$\Phi_0$ is arbitrary. However since $\Phi_0$ is the wave function for a bound
state it can be normalized. In the present context we choose 
$\langle \Phi_0 | P\Psi \rangle = 1$. 

\section{Corrections to the mean field energy}
\label{sec:corrections}

The corrections to the shell model energy $\bar E_0$ 
are obtained in the statistical formalism as in the case 
of the nuclear reactions \cite{Fes92,Fes80},  by expanding (\ref{deltae})
in a series of contributions which, rather than by powers of a coupling 
constant, are identified by the complexity of the states of the $Q$-space.

Thus one writes
\begin{equation}
\Delta E = \sum_{m =1}^r \Delta E_m \, ,
\label{deltaeexp}
\end{equation}
where
\begin{eqnarray}
\Delta E_m & = &  \langle \Phi_0 |\, V_{01} G_1 V_{12} G_2 
\cdots V_{m-1 , m} G_m V_{m,m-1} G_{m-1}
\cdots V_{10}\, | \Phi_0 \rangle \, ,
\label{deltaem} \\
V_{ij} & = &  Q_i V Q_j \, ,
\label{vij}
\end{eqnarray}
and $Q_j$ is the operator projecting into the $j$ particles--$j$ holes sector 
of the $Q$-space.

\begin{figure}[tb]
\begin{center}
\mbox{\epsfig{file=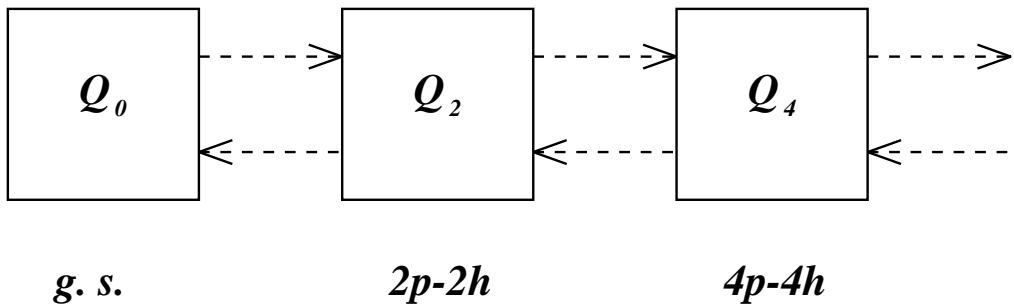}}
\vskip 3mm
\caption{
The partition of the Hilbert space of nuclear matter in sets of increasing 
complexity. The first box on the left defines the $P$-space, the second one 
embodies the simplest states in the $Q$-space and so on.}
\label{fig:boxes}
\end{center}
\end{figure}

One thus sees that the states belonging to the $Q$-space are 
classified into $r$ sets, each set incorporating a well-defined
type of excitation; specifically the $k$-th set embodies all the 
$k$ particles--$k$ holes states. 
Actually, because of the essentially two-body nature of the residual interaction
$V$, the one effectively responsible for the passage from one set of states to
the other, only excitations corresponding to an {\em even} number of particles
(and holes) are actually filling the sets, as indicated in Fig.~\ref{fig:boxes}.
The propagation of the system into the $k$-th set is then described by the 
operator $G_k$, which when $k =r$ reads
\begin{equation}
G_r = \frac{1}{E - h_{rr}} \  \  ,
\label{grfin}
\end{equation}
whereas for $k < r$ obeys the recurrence relation
\begin{equation}
G_k = \frac{1}{E - h_{kk} - V_{k,k+1} G_{k+1} V_{k+1,k}} \  \  .
\label{gk}
\end{equation}
In the above 
\begin{equation}
h_{kk} = H_{Q_k  Q_k} + V_{Q_k P}\, \left (\frac{1}{E - \bar {\cal H}}
\right )^\prime\, V_{P Q_k}  \, .
\label{hkkdef}
\end{equation}
 For the sake of illustration each set is sometimes displayed as a box: The 
order among the sets, which is a condition for the validity of (\ref{deltaem}),
can then be expressed by stating that no box can be occupied unless the previous
one has already been too (see Fig.~\ref{fig:boxes}).

Moreover, it should be noticed that in a nucleus the expansion
(\ref{deltaeexp}) is finite since the excited states with the greatest
complexity correspond to the case $r =A$, $A$ being the mass number.

Since the mean field contains the average, the average of the corrections to
the mean field vanishes. In addition because of the random nature of the
processes occurring beyond the mean field the average of the corrections to the
shell model stemming from any individual set vanishes as well. Thus 
\begin{equation}
\langle \Delta E \rangle = \langle \Delta E_1 \rangle + 
\langle \Delta E_2 \rangle + \cdots +
\langle \Delta E_A \rangle = 0 
\label{avdeltae}
\end{equation}
and 
\begin{equation}
\langle \Delta E_n \rangle = 0 \, .
\label{avdeltaen}
\end{equation}
However the average of the square of the corrections does not vanish. 
 For this,
namely for 
\begin{equation}
\langle (\Delta E)^2 \rangle =  \langle\, 
\sum_{n,m} \Delta E_n \,  \Delta E_m \, \rangle \, ,
\label{deltae2av}
\end{equation}
the randomness of the $Q$-space physics is exhibited in the randomness of the 
nuclear matrix elements entering into the definition of $\Delta E_n$.
As a consequence, the off-diagonal elements in (\ref{deltae2av}) disappear, 
leaving
\begin{equation}
\langle (\Delta E )^2 \rangle =  \langle\,
\sum_{n}(\Delta E_n )^2\,  \rangle \, .
\label{deltae2avf}
\end{equation}

The implementation of the requirement
\begin{equation}
\langle \Delta E_1 \rangle = \langle \Delta E_2 \rangle = \cdots
= \langle \Delta E_A \rangle =0
\label{deltaei0}
\end{equation}
is achieved, in  the case $r = 1$,  by redefining $\Delta E_1$ itself according 
to 
\begin{equation}
 \Delta E_1  = \langle\, \phi_A^0\, |\, V_{01} G_1  V_{10}\,
 |\, \phi_A^0\, \rangle - 
\left [ \langle\, \phi_A^0\, |\, V_{01} G_1  V_{10}\,
 |\, \phi_A^0\, \rangle \right ]_{AV} \, ,
\label{avdeltae1}
\end{equation} 
and likewise for the terms with $r \not= 1$.
In (\ref{avdeltae1}) $\phi_A^0$ represents the nuclear mean field ground state 
wave function and the 
square brackets mean energy averaging.

Now, the corrections to the mean field can be calculated, 
according to the formula (\ref{deltae}), provided the eigenfunctions 
$\psi_{k,\alpha}$ and the eigenvalues $\epsilon_{k,\alpha}$ of the operator 
$G^{-1}_k$ are known, which in turns implies that the equation
\begin{equation}
  G^{-1}_k \psi_{k,\alpha} = (E-\epsilon_{k,\alpha}) \psi_{k,\alpha}
\end{equation}
has to be solved for any value of the index $k$.

To begin with we consider $\Delta E_1$:
\begin{equation}
  \Delta E_1 
  = \sum_\beta\frac{|\langle\psi_{2\beta}|V|\phi_A^0\rangle|^2}
    {E-\epsilon_{2\beta}}
    - \left[\sum_\beta
    \frac{|\langle\psi_{2\beta}|V|\phi_A^0\rangle|^2}
    {E-\epsilon_{2\beta}}\right]_{AV}
    \, ,
\end{equation}
where $\epsilon_{2\beta}$ and $|\psi_{2\beta}\rangle$ are the eigenvalues and
eigenvectors, respectively, corresponding to the 2p--2h excitations.

However, as previously remarked, in the present scheme only the square of the
corrections to the mean field, namely
\begin{equation}
  (\delta E_1)^2 = \left[ (\Delta E_1)^2 \right]_{AV} - \left[ (\Delta E_1)
    \right]_{AV}^2 \, ,
\label{eq:deltaE1}
\end{equation}
are meaningful. We are left with the task of
evaluating 
\begin{equation}
  (\delta E_1)^2 = \left[ \sum_{\beta\gamma}
  \frac{|\langle\psi_{2\beta}|V|\phi_A^0\rangle|^2 
       |\langle\psi_{2\gamma}|V|\phi_A^0\rangle|^2}
       {(E-\epsilon_{2\beta})(E-\epsilon_{2\gamma})} \right]_{AV} 
  - \left[ \sum_{\beta}\frac{|\langle\psi_{2\beta}|V|\phi_A^0\rangle|^2}
                           {E-\epsilon_{2\beta}} \right]^2_{AV} \, .
\label{eq:deltaE1bis}
\end{equation}
 For the first term of the above equation we have
\begin{eqnarray}
  &&  
  \left[ \sum_{\alpha, \beta}
    \frac{\langle \phi_A^0
    |V|\psi_{2\alpha}\rangle\langle\psi_{2\alpha}|V|\phi_A^0\rangle 
    \langle \phi_A^0|V|\psi_{2\beta}\rangle\langle\psi_{2\beta}|V|\phi_A^0
     \rangle}
         {(E-\epsilon_{2\alpha})(E-\epsilon_{2\beta})} \right]_{AV}
    \nonumber \\
  && \cong 
  \left( \frac{1}   {E-\bar \epsilon_{2}} \right)^2
\left[ \sum_{\alpha, \beta}
    \langle \phi_A^0
    |V|\psi_{2\alpha}\rangle\langle\psi_{2\alpha}|V|\phi_A^0\rangle 
    \langle \phi_A^0|V|\psi_{2\beta}\rangle\langle\psi_{2\beta}|V|\phi_A^0
     \rangle \right]_{AV} \, 
\end{eqnarray}
where $\bar \epsilon_2$ is the average excitation for the 2p--2h states.
To obtain the random phase average of the quantity in the square brackets
we make use of the result:
\begin{equation}
 \langle A A^* B B^* \rangle = \langle A A^* \rangle
\langle B B^* \rangle  + \langle A^* B \rangle
 \langle A B^* \rangle + \text{quartic terms} \, .
 \label{aveab}  
\end{equation}
We shall neglect quartic terms.

The first term $\langle A A^* \rangle
\langle B B^* \rangle$ cancels exactly the second term in 
eq.~(\ref{eq:deltaE1bis}). Evaluating the second term in eq.~(\ref{aveab}) 
yields 
\begin{eqnarray}
  \left[ \left( \delta E_1 \right)^2 \right] &=& 
 \left( \frac{1}{E - \bar \epsilon_2}\right)^2 \sum_{\alpha}
    \left[ |\langle \phi_A^0
    |V|\psi_{2\alpha}\rangle|^2_{AV} \right]^2
    \nonumber \\
\label{avedeltae11}
  & \cong &
  \frac{\overline {\Delta E_1}}{(E - \bar \epsilon_2)^2}\frac{1}{D_2}
    \left[ |\langle \phi_A^0
    |V|\psi_{2}\rangle|^2_{AV} \right]^2 \, . 
\label{avedeltae1}
\end{eqnarray}
In this last equation $\overline {\Delta E_1}$ 
is the interval over which the energy
average is carried out. It is essentially equal to $\epsilon$. 
The quantity $D_2$ is the energy distance between two neighboring 2p--2h
states. Finally $\psi_2$ is meant as a representative of the 2p--2h excitations.

We now go on to consider $\Delta E_2$: 
\begin{eqnarray}
\Delta E_2 &=& 
 \langle \phi_A^0 |\, V_{01} G_1 V_{12} G_2 V_{21} G_{1}
V_{10}\, | \phi_A^0 \rangle  
\nonumber \\  
&=& \sum_{\alpha, \beta, \alpha^\prime} 
\frac{\langle \phi_A^0
    |V|\psi_{2\alpha}\rangle}{E-\epsilon_{2\alpha}}
\frac{\langle \psi_{2\alpha}
    |V|\psi_{4\beta}\rangle}{E-\epsilon_{4\beta}}
\frac{\langle \psi_{4\beta}
    |V|\psi_{2\alpha^\prime}\rangle}{E-\epsilon_{2\alpha^\prime}}
 \langle \psi_{2\alpha^\prime}
    |V|\phi_A^0\rangle
    \nonumber \\
&\cong& \frac{1}{(E-\bar\epsilon_2)^2} \frac{1}{(E-\bar\epsilon_4)}  
\left[ \sum_{\alpha, \beta} \langle \phi_A^0
    |V|\psi_{2\alpha}\rangle
\langle \psi_{2\alpha}
    |V|\phi_A^0\rangle \right.
\nonumber \\
&& \cdot \langle \psi_{2\alpha}
    |V|\psi_{4\beta}\rangle \langle \psi_{4\beta}
    |V|\psi_{2\alpha}\rangle
 \nonumber \\
&& \left.   - \left[ \sum_{\alpha, \beta} |\langle \phi_A^0
    |V|\psi_{2\alpha}\rangle|^2
|\langle \psi_{2\alpha}
    |V|\psi_{4\beta}\rangle|^2 \right]_{AV} \right]
\end{eqnarray}
where we have set $\alpha = \alpha^\prime$ and in the
last line have  subtracted the average of $\Delta E_2$ as written in the first
line so that $\langle \Delta E_2 \rangle = 0$. We now consider the square of the modified
$\Delta E_2 = \delta E_2$. We obtain
\begin{eqnarray}
(\delta E_2)^2  &\cong& 
\frac{1}{(E- \bar \epsilon_2)^4}\frac{1}{(E- \bar \epsilon_4)^2}\left[
 \sum_{\alpha, \beta, \alpha^\prime, \beta^\prime}
\langle \phi_A^0
    |V|\psi_{2\alpha}\rangle
\langle \psi_{2\alpha}
    |V|\phi_A^0\rangle \right.
\nonumber \\
&& \quad 
 \langle \psi_{2\alpha}
    |V|\psi_{4\beta}\rangle  \langle \psi_{4\beta}
    |V|\psi_{2\alpha}\rangle 
\bigg\{  \langle \phi_A^0
    |V|\psi_{2\alpha^\prime}\rangle  \langle \psi_{2\alpha^\prime}
    |V|\phi_A^0\rangle  
    \nonumber \\
&& \quad   \langle \psi_{2\alpha^\prime}
    |V|\psi_{4\beta^\prime}\rangle   \langle \psi_{4\beta^\prime}
    |V|\psi_{2\alpha^\prime}\rangle \bigg\} -  
 \sum_{\alpha, \beta, \alpha^\prime, \beta^\prime}
\left[ |\langle \phi_A^0
    |V|\psi_{2\alpha}\rangle |^2 \right.
\nonumber \\
&& \quad \left. \left.  |\langle \phi_A^0
    |V|\psi_{2\alpha^\prime}\rangle |^2
|\langle \psi_{2\alpha}
    |V|\psi_{4\beta}\rangle |^2 |\langle \psi_{2\alpha^\prime}
    |V|\psi_{4\beta^\prime}\rangle |^2 \right]_{AV} \bigg]_{AV} \right. \, .
\end{eqnarray}
To calculate the average we now invoke eq.~(\ref{aveab}). The first term again
cancels the subtracted term and one is left with 
\begin{equation}
(\delta E_2)^2 =
\frac{1}{(E- \bar \epsilon_2)^4}\frac{1}{(E- \bar \epsilon_4)^2}\left[
 \sum_{\alpha, \beta}
|\langle \phi_A^0
    |V|\psi_{2\alpha}\rangle|^4
|\langle \psi_{2\alpha}
    |V|\psi_{4\beta}\rangle|^4 \right]_{AV} \, .
\label{eqdeltae2int}
\end{equation}
Comparing this result with that for
$(\delta E_1)^2$, eq.~(\ref{avedeltae1}), one finds
\begin{equation}
(\delta E_2)^2 =
\frac{1}{(E- \bar \epsilon_2)^2}\frac{1}{(E- \bar \epsilon_4)^2}(\delta E_1)^2
|\langle \psi_{2}
    |V|\psi_{4}\rangle|^4
\label{eqdeltae2d1}
\end{equation}
which gives us an expression for the {\em expansion parameter}, though, as it
has been emphasized earlier, the expansion has a finite number of terms.
Again in the above $\psi_4$ is portraying a representative of the 4p--4h states. 
Eq.~(\ref{eqdeltae2d1}) can be generalized into a relation between 
$(\delta E_n)^2$ and $(\delta E_{n-1})^2$:
\begin{equation}
\left[ (\delta E_n)^2\right]  =
\frac{1}{(E- \bar \epsilon_{2n})^2}\frac{1}{(E- \bar \epsilon_{2n-2})^2}
\left[ (\delta E_{n-1})^2 \right] 
\left[ |\langle \psi_{2n-2}
    |V|\psi_{2n}\rangle|^4\right]  \, .
\label{eqdeltae2dn}
\end{equation}

This concludes the general development of the statistical theory. To make
further progress a detailed description of the system under 
consideration is 
required. Our application  will be to nuclear matter. But one should bear in
mind that up to this point our results can be applied to any system to validate
the mean field concept.

\section{The projection operator and the HF theory}
\label{sec:HF}

To obtain the true energy ($E$) and the {\em mean field} one ($\bar E_0$)
 (the shell model energy for a nucleus) 
 we now need an explicit form for both the projection operator $P$
and the NN interaction.

Let us start first by searching for a suitable projection operator. 
Suppose, for this purpose, that for a given interaction one has solved 
the associated HF problem. Let us call $\{ \phi_{\text{HF}}^i \}$, with the 
index $i = 1,2, \dots$, the single particle wave functions of the HF 
orbitals, which form an orthonormal complete set for the Hilbert space 
of a one particle system. Out of the $\{ \phi_{\text{HF}}^i \}$ we can build an 
infinite set of Slater determinants $\{ \chi_{\text{HF}}^i \}$  which also form 
an orthonormal complete set for the Hilbert space of the nucleus.

We can then define 
\begin{equation}
P = \sum_{i=1}^M |\ \chi_{\text{HF}}^i\ \rangle\langle\  \chi_{\text{HF}}^i\ | 
\label{pprojd}
\end{equation}
which, in the special case of $M=1$, reduces to 
\begin{equation}
P = |\ \chi_{\text{HF}}^1\ \rangle\langle\  \chi_{\text{HF}}^1\ |  \, ,
\label{pproj1}
\end{equation}
$| \chi_{\text{HF}}^1 \rangle$ being the ground state HF determinant of the 
nuclear system. 
In the following we shall stick to the option $M=1$,  although generalizations 
to larger values of $M$, while cumbersome, should be useful to consider. 

Notice that, with $P$ given by (\ref{pproj1}), the wave functions $(P\Psi )$ 
and $\langle P\Psi  \rangle$  turn out to be proportional to each other 
and to the ground state HF determinant $| \chi_{\text{HF}}^1 \rangle$ as they 
must since the $P$ space has only one member.

With this projection operator and the mean field given by $\bar {\cal H}$
 (eq.~(\ref{hbarv}))
 \begin{equation}
  {\bar E_0} = E_{\text{HF}} +
\langle\ \chi_{\text{HF}}^1\ |  VQQV^\dagger 
|\ \chi_{\text{HF}}^1\ \rangle 
\frac{1}{\bar E_0 -E -\epsilon}
\label{e0baravex}
\end{equation}
 where
\begin{equation}
E_{\text{HF}} = 
\langle\ \chi_{\text{HF}}^1\ |\, H_{PP}\, |\ \chi_{\text{HF}}^1\ \rangle \, . 
\label{ehfdef}
\end{equation}
and
\begin{equation}
  V = H \sqrt{\frac{\bar{E}_0-\epsilon-E}{\bar{E}_0-\epsilon-H_{QQ}}}
\label{eq:eff-int}
\end{equation}
(see eq.~(\ref{vpqdefr})).
Eq.~(\ref{e0baravex}), writing $Q=1-P$, becomes
\begin{equation}
\bar E_0 = E_{\text{HF}} + 
\frac{1}{\bar E_0 - E -\epsilon} \, 
\bigg\{ \langle\ \chi_{\text{HF}}^1\ |\, V V^\dagger\, 
|\ \chi_{\text{HF}}^1\ \rangle\, -  \vert \langle\ \chi_{\text{HF}}^1\ |\, V\,
|\ \chi_{\text{HF}}^1\ \rangle\vert^2 \bigg\} \, .
\label{e0barfin}
\end{equation}

The expression (\ref{e0barfin}) is general. 
 From now on, however, we shall confine ourselves to consider a Fermi liquid
(infinite nuclear matter).
 For such a system $|\chi_{\text{HF}}^1 \rangle = | F \rangle$, $| F \rangle$ 
being the wave function of a Fermi sphere of radius $k_F$ (the Fermi 
wavenumber).

In infinite nuclear matter the quantity in curly brackets in 
eq.~(\ref{e0barfin}), which we refer to as the {\em variance} of the residual 
interaction $V$, can be simplified. One has (the states 
$| n \rangle$ labelling the spectrum of the Fermi sphere)
\begin{eqnarray}
\langle F |\, V V^\dagger\, | F  \rangle & = &
\sum_n \langle F |\, V\, | n \rangle \langle n |\, V^\dagger\, | F  
\rangle \nonumber\\
 & \cong & \langle F |\, V\, | F  \rangle \langle F |\, V^\dagger\, | F \rangle
\nonumber\\
&  & + \sum_{1p-1h} \langle F |\, V\, | 1p-1h \rangle \langle
1p-1h  |\, V^\dagger\, | F \rangle \label{fv2fsv}\\
&  & +  \sum_{2p-2h} \langle F |\, V\, | 2p-2h \rangle \langle 
2p-2h  |\, V^\dagger\, | F \rangle  \nonumber\\
& = & \vert \langle F |\, V\, | F  \rangle \vert ^2 
+ \beta^2  \, , \nonumber
\end{eqnarray}
the piece related to the 1p-1h states essentially vanishing as required by the 
Brillouin theorem \cite{Gro91} and further terms in the right hand side being 
neglected because of the prevalent two-body character of the residual 
interaction $V$.
The quantity $\beta^2$ is just the value of the variance given in 
eq.~(\ref{e0barfin}). 
The latter is of course hard to deal with because it is not only energy
dependent, but dependent upon the energy averaging procedure as well.
  Furthermore, its connection to the $Q$-space Hamiltonian is highly non-linear.
However, in nuclear matter the variance of the residual interaction is 
{\em positive} since $\beta^2$ (defined by the eq.~(\ref{fv2fsv}) itself) is 
positive.
Moreover, by inserting (\ref{fv2fsv}) into (\ref{e0barfin}), one obtains the 
important expression
\begin{equation}
\bar E_0 \cong E_{\text{HF}} + 
\frac{\beta^2}{\bar E_0 - E -\epsilon} \, ,
\label{e0barfin2}
\end{equation}
which embodies all the energies characterizing our problem, namely the HF, the 
mean field (which would correspond to the shell model in a finite nucleus) and 
the true one.

Thus, even if a residual interaction $V$ is given and the associated HF energy
$E_{\text{HF}}$ is calculated, still the energy $E$ of the system cannot be 
obtained since the mean field energy $\bar E_0$ remains to be fixed. Yet, if 
as a first orientation one sets $\bar E_0 \simeq E$, then the remarkable result
\begin{equation}
\bar E_0  \cong E_{\text{HF}} - \frac{\beta^2}{\epsilon} 
\label{e0barappr}
\end{equation}
follows, showing that the mean field energy is {\em lower} than the HF 
one, the parameter $\epsilon$ being of course positive. 
This result holds only if $|\bar{E}_0-E| \ll \epsilon$.

\section{Spectroscopic factor}
\label{sec:spectr-fact}

The projected wave function $P\Psi$ is just a component of the full wave
function $\Psi$. Just how big a fraction of $\Psi$ is contained in $P\Psi$ is
obviously of interest. Toward this end let 
$\langle P\Psi|P\Psi\rangle = S^2$. We will now obtain an equation for $S^2$ in
terms of the other quantities characterizing our problem. We begin with the
identity 
\begin{equation}
  S^2 = 1 - \langle Q\Psi|Q\Psi\rangle \, .
\end{equation}
Using eq.~(\ref{qpsiphi0}) for $(Q\Psi)$, $S^2$ becomes
\begin{equation}
  S^2 = 1 - \langle \Phi_0|V_{PQ}\left(\frac{1}{E-h_{QQ}}\right)^2 V_{QP}
    |\Phi_0\rangle \, .
\end{equation}
We now take the Fermi gas model to be $\Psi$, written $|F\rangle$.
Then $\Phi_0=S|F\rangle$ and
\begin{equation}
  S^2 = 1 - S^2 \langle F|V_{PQ}\left(\frac{1}{E-h_{QQ}}\right)^2 V_{QP}
    |F\rangle \ .
\end{equation}
Now, a little algebra allows us to recast the previous equation into the form
\begin{eqnarray}
  S^2 &\cong& 1 + S^2 \langle F|V_{PQ}\frac{d\phantom{E}}{dE}
    \left(\frac{1}{E-h_{QQ}}\right) V_{QP}|F\rangle \nonumber \\
  &=& 1 + S^2 \left[ \langle F|\frac{d\phantom{E}}{dE}\left(
    V_{PQ}\frac{1}{E-h_{QQ}}V_{QP}\right)|F\rangle \right. \nonumber \\
  &&\qquad - \left. \langle F|\frac{dV_{PQ}}{dE}\frac{1}{E-h_{QQ}}V_{QP}
    |F\rangle - \langle F|V_{PQ}\frac{1}{E-h_{QQ}}\frac{dV_{QP}}{dE}|F\rangle
    \right] \, ,
\end{eqnarray}
where the weak energy dependence of $h_{QQ}$ has been neglected.
Moreover, since
\begin{mathletters}
\begin{equation}
  \frac{dV_{PQ}}{dE} = -V_{PQ}\frac{1}{2(\bar{E}_0-\epsilon-E)}
\end{equation}
and
\begin{equation}
  \frac{dV_{QP}}{dE} = -V_{QP}\frac{1}{2(\bar{E}_0-\epsilon-E)}
\end{equation}
\end{mathletters}
it follows that 
\begin{equation}
  S^2 = 1 + S^2 \left[ \frac{d\phantom{E}}{dE}\langle F|V_{PQ}\frac{1}{E-h_{QQ}}
    V_{QP}|F\rangle + \frac{1}{\bar{E}_0-\epsilon-E}
    \langle F|V_{PQ}\frac{1}{E-h_{QQ}} V_{QP}|F\rangle \right]
\end{equation}
or, by virtue of (\ref{deltae}), that
\begin{eqnarray}
  S^2 &=& 1 + S^2\left[\frac{d\phantom{E}}{dE}(E-\bar{E}_0) +
    \frac{E-\bar{E}_0}{\bar{E}_0-\epsilon-E}\right] \nonumber \\
  &=& 1 - S^2 \left( \frac{d\bar{E}_0}{dE} +
  \frac{\epsilon}{\bar{E}_0-\epsilon-E} \right) \, .
\label{eq:S2Ebar0}
\end{eqnarray}
The explicit dependence of $\bar{E}_0$ upon $E$ is provided by (\ref{e0barfin2})
which, when inverted, yields
\begin{equation}
  \bar{E}_0 = \frac{1}{2}\left[E+\epsilon+E_{\text{HF}}\pm
    \sqrt{(E+\epsilon-E_{\text{HF}})^2 + 4\beta^2}\right] \, ,
\end{equation}
where the minus sign in front of the square root should be taken, since for 
vanishing residual interaction ($\beta^2=0$) the mean field should reduce to 
the HF field ($\bar{E}_0=E_{\text{HF}}$). Hence
\begin{equation}
  \frac{d\bar{E}_0}{dE} = \frac{1}{2} + \frac{1}{2} 
    \frac{E_{\text{HF}}-E-\epsilon-2\,d\beta^2/dE}
         {\sqrt{(E_{\text{HF}}-E-\epsilon)^2+4\beta^2}} \, ,
\end{equation}
which, inserted into (\ref{eq:S2Ebar0}), finally leads to
\begin{eqnarray}
  S^2 &=& \left[ \frac{3}{2} + \frac{1}{2}
    \frac{E_{\text{HF}}-E-\epsilon-2\,d\beta^2/dE}
         {\sqrt{(E_{\text{HF}}-E-\epsilon)^2+4\beta^2}}
    +\frac{\epsilon}{\bar{E}_0-E-\epsilon}\right]^{-1} \nonumber \\
  &=& \left[ \frac{3}{2} + \frac{1}{2}
    \frac{E_{\text{HF}}-E-\epsilon-2\,d\beta^2/dE}
         {\sqrt{(E_{\text{HF}}-E-\epsilon)^2+4\beta^2}}
    + \epsilon\frac{\bar{E}_0-E_{\text{HF}}}{\beta^2}\right]^{-1} \, ,
\label{eq:S2constr}
\end{eqnarray}
namely to the equation we were looking for.
Eq.~(\ref{eq:S2constr}) expresses the spectroscopic factor $S^2$ in terms of the
HF, mean field and true energies, with an additional dependence upon the energy
averaging parameter $\epsilon$, the residual interaction $\beta^2$ and the
derivative of the latter with respect to the energy.
Since all these quantities are either explicitly evaluated ($E_{\text{HF}}$,
$\bar{E}_0$ and $E$) or fixed, in principle, both by the experiment ($\epsilon$)
and by our theoretical framework ($\beta^2$), eq.~(\ref{eq:S2constr}) provides
an important check for the consistency of our approach, also because
independent estimates on the depletion of the ground state wave 
function are available for lead, a nucleus to which the nuclear matter concepts 
should apply.

\section{ A schematic model}
\label{sec:numerical-results}

In this section we shall employ a schematic model to gain an insight into the
effectiveness and self-consistency of the formalism described above.
The model will rely on two equations which, for sake of convenience, 
are repeated
here. The first of these is 
eq.~(\ref{e0barfin2})

\begin{eqnarray}
&&\bar E_0 \cong E_{\text{HF}} + 
\frac{\beta^2}{\bar E_0 - E -\epsilon} \, .
\nonumber
\end{eqnarray}
The second is obtained from Eq.~(\ref{avedeltae11}), i.~e. 
\begin{eqnarray}
  \left[ \left( \delta E_1 \right)^2 \right] = 
 \left( \frac{1}{E - \bar \epsilon_2}\right)^2 \sum_{\gamma}
    \left[ |\langle \phi_A^0
    |V|\psi_{2\gamma}\rangle|^2_{AV} \right]^2 \, , 
 \nonumber
\end{eqnarray}
which is approximated by 
\begin{equation}
 \left[ (\delta E_1)^2 \right]  = \frac{\beta^4}{(E-\alpha\epsilon)^2} \, ,
 \label{eqst1}
\end{equation}
where $\alpha$ is a parameter which estimates the average energy of the 2p--2h
states involved (we set $\bar \epsilon_2 = \alpha \epsilon$ for convenience) 
and {\em importantly} helps to correct for the poor
approximation of $\beta^4$ for the sum over the 2p--2h states $\gamma$ in 
eq.~(\ref{avedeltae11}).
Taking the square root of eq.~(\ref{eqst1}) yields
\begin{equation}
  E-\bar{E}_0 = \pm \frac{\beta^2}{E-\alpha\epsilon} \, .
\label{eq:fluct-guess}
\end{equation}

If now one combines eq.~(\ref{e0barfin2}) and (\ref{eq:fluct-guess}), using the upper
sign (plus) one obtains a lower bound $E_l$ for the energy of nuclear matter, 
while using the lower sign (minus) one obtains an upper bound $E_u$.
Moreover, one obtains differing values of the mean field energy to be denoted
by $\bar{E}_0^l$ and $\bar{E}_0^u$, respectively. We ask whether there is a
choice of $\epsilon$, $\alpha$ and $\beta$ such that 
\begin{itemize}
\item[i)] the two mean field energies $\bar{E}_0^{l}$ and $\bar{E}_0^{u}$ 
turn out to be the same over a 
range of densities (or of Fermi momenta $k_F$) of significance for nuclear 
matter;
\item[ii)] the ``experimental'' values of the binding energy, saturation density
and compression modulus of nuclear matter, namely $B.E./A=-16$~MeV, 
$k_F=1.36$~fm$^{-1}$ and $K=23$ or 16 MeV (the two values refer to a hard and a 
soft equation of state, respectively) \cite{Bet71} are accounted for in a sense to be later 
specified;
\item[iii)] the spectroscopic factor given by eq.~(\ref{eq:S2constr}) turns out
to be meaningful, in the sense of being of course less than one both on the 
lower bound (where its value $S_l$ is associated with $E_l$) 
and on the upper one (where its value $S_u$ is associated with $E_u$).
 Furthermore $S_l$ and $S_u$ should not be too much at variance with existing 
estimates.
\end{itemize}

To satisfy condition i) we proceed by first eliminating $\bar{E}_0^u$ and
$\bar{E}_0^l$ from eq.~(\ref{e0barfin2}) and (\ref{eq:fluct-guess}).
This yields an equation for $\beta_l^2$ and $\beta_u^2$:
\begin{mathletters}
\label{eq:beta2sol}
\begin{equation}
  \beta^2_l = \frac{E_l-\alpha\epsilon}{2}\left\{[2E_l-\epsilon(\alpha+1)
    -E_{\text{HF}}]-\sqrt{[2E_l-\epsilon(\alpha+1)-E_{\text{HF}}]^2 
    +4\epsilon(E_l-E_{\text{HF}})}\right\} \, ,
\end{equation}
for the lower bound, and
\begin{eqnarray}
  &&\beta^2_u = 
    \frac{E_u-\alpha\epsilon}{2}\left\{[E_{\text{HF}}-\epsilon(\alpha-1)]
    -\sqrt{[E_{\text{HF}}-\epsilon(\alpha-1)]^2 
    +4\epsilon(E_u-E_{\text{HF}})}\right\} \, ,
\end{eqnarray}
\end{mathletters}
for the upper bound.
Note that the solutions with the plus sign in front of the square root have been
discarded, since they lead to an incorrect limit as $\epsilon \to 0$.
The energy derivatives of (\ref{eq:beta2sol}) read then
\begin{mathletters}
\begin{eqnarray}
  \frac{d\beta^2_l}{dE} &=& \frac{1}{2}\left\{[2E_l-\epsilon(\alpha+1)
    -E_{\text{HF}}]-\sqrt{[2E_l-\epsilon(\alpha+1)-E_{\text{HF}}]^2 
    +4\epsilon(E_l-E_{\text{HF}})}\right\} \nonumber \\
  && \quad+(E_l-\alpha\epsilon)\left\{1-\frac{2E_l-\alpha\epsilon-E_{\text{HF}}}
    {\sqrt{[2E_l-\epsilon(\alpha+1)-E_{\text{HF}}]^2+
    4\epsilon(E_l-E_{\text{HF}})}}
    \right\} \, ,
\end{eqnarray}
for the lower bound, and
\begin{eqnarray}
  \frac{d\beta^2_u}{dE} &=& \frac{1}{2}\left\{[E_{\text{HF}}-\epsilon(\alpha-1)]
    -\sqrt{[E_{\text{HF}}-\epsilon(\alpha-1)]^2 
    +4\epsilon(E_u-E_{\text{HF}})}\right\} \nonumber \\
  && \quad - (E-\alpha\epsilon)\frac{\epsilon}
    {\sqrt{[E_{\text{HF}}-\epsilon(\alpha-1)]^2+4\epsilon(E_u-E_{\text{HF}})}}
    \, ,
\end{eqnarray}
\end{mathletters}
for the upper bound, which, when inserted into (\ref{eq:S2constr}), provide an
expression for the spectroscopic factor, associated with the lower and the 
upper bound, respectively, in terms of $E_l$ ($E_u$), $\bar{E}_0^l$ 
($\bar{E}_0^u$), $E_{\text{HF}}$, $\alpha$ and $\epsilon$.

 From eq.~(\ref{eq:beta2sol}) we note that inserting values of $\bar{E}_l$ (or
$\bar{E}_u$) and of $E_{\text{HF}}$ one obtains a value of $\beta_l^2$ and
$\beta_u^2$ for given values of $\alpha$ and $\epsilon$.
These values of $\beta^2$ can then be inserted into eq.~(\ref{e0barfin2}) to obtain
$\bar{E}_0^l$ and $\bar{E}_0^u$. A solution of eq.~(\ref{e0barfin2}) and
(\ref{eq:fluct-guess}) occurs when these upper and lower bounds for the mean field
are equal.

But how are $E_l$ and $E_u$ chosen? And how is the value of $E_{\text{HF}}$ to
be obtained? The first of these questions is answered as follows:
\begin{mathletters}
\begin{eqnarray}
E_l &=& {\cal E}-W/2 \\
E_u &=& {\cal E}+W/2 \, ,
\end{eqnarray}
\end{mathletters}
where
\begin{equation}
  {\cal E} = [-16 +39.5(k_F-1.36)^2]\, \text{MeV} 
\label{eq:enuclmatt}
\end{equation}
incorporates the present knowledge on the ground state energy of nuclear matter,
quoted in ii), as extrapolated from finite nuclei. $W$ is the fluctuation energy.

The value of $E_{\text{HF}}$ as a function of $k_F$ is obtained from an assumed
two-body interaction. This is described in Appendix~\ref{app:app}.
The resulting Hartree-Fock energy for a Fermi gas is shown in Fig.~\ref{fig:HF}
and, for a reduced scale in $k_F$, in Fig.~\ref{fig:bind-energy}.
 From Fig.~\ref{fig:HF} we see that for the assumed two-body interaction a
minimum of the binding energy as a function of $k_F$ occurs at $k_F=1.78$ 
fm$^{-1}$ with the corresponding energy per particle of $-$7.3 MeV.

The potential can also be used to calculate the bare value of $\beta^2$. 
This is discussed in 
Appendix~\ref{app:appb}. Obtained values are given in Tables
\ref{tab:beta2_2}, \ref{tab:beta2_3} and \ref{tab:beta2_4}:
they range from $4.5\times10^4$ to $7.0\times10^4$ MeV$^2$/nucleon.

In discussing our findings we first observe that, for a given fluctuation energy
$W$, a whole set of values for the parameters $\alpha$ and $\epsilon$ 
exist such to satisfy the requirement i), namely 
$\bar{E}_0^l =\bar{E}_0^u$, at $k_F =1.36$ fm$^{-1}$.
They lie on the curves displayed in Fig.~\ref{fig:alfeps}. 
 Furthermore $\alpha$ and $\epsilon$ should also be such to fulfill the constraint
\begin{equation}
\bar \epsilon_2 = \alpha \epsilon \ge 20 {\text { MeV}} \, ,
\end{equation}
which represents a fair estimate of the lower limit 
for the excitation energy of the 2p--2h states.
This restricts the acceptable values for $\alpha$ and $\epsilon$ 
to the domain to the right of the dotted line in 
 Fig.~\ref{fig:alfeps}.

In this region $\alpha$ and $\epsilon$  should be selected in such a way to
comply with the requirements ii) and iii).
This turned out to be possible and our results are shown in 
 Fig.~\ref{fig:bind-energy} and Tables \ref{tab:beta2_2}, \ref{tab:beta2_3} and
\ref{tab:beta2_4}.

However there one sees that as one moves away from $k_F =1.36$ fm$^{-1}$
the equality between $\bar{E}_0^l$ and  $\bar{E}_0^u$
is no longer exactly satisfied, although the two mean fields remain rather close
in the range $0.9 \le k_F \le 1.5$  fm$^{-1}$
providing the values of $W$ are ``moderate''.
Indeed, quantitatively, requirement i) can be reasonably satisfied in the above
quoted range of $k_F$ for $W \le 4$ MeV only. For larger fluctuation energies 
$\bar{E}_0^l$ and  $\bar{E}_0^u$ differ too much: an orientation on the size of
the error around the mean field in thus obtained.

Concerning the energy averaging parameter, $\epsilon$, is 1 MeV larger that $W$ for all
the cases listed in the tables. The values of the spectroscopic factors 
$S_l$ and $S_u$ are quite stable, while the values of the effective interaction
$\beta_l^2$ and $\beta_u^2$ grow rapidly with increasing $W$.
Most importantly, their values differ by three orders of magnitude from the bare
$\beta^2$. This is in part due to the renormalization (see eqs.~(\ref{vpqdefr})
and (\ref{vqpdefr})) induced by the energy averaging.
The random phase averaging also gives rise to a reduction in the magnitude of
the residual interaction. Rough calculations indicate that these two effects
are sufficient to produce the observed sharp reduction.

\section{Concluding Remarks}
\label{sec:conclusions}

The analysis presented in sections \ref{sec:intro}--\ref{sec:corrections}
is based on two propositions. It is assumed that the mean field is the slowly
varying component of the nuclear interaction, which can be obtained by taking an
appropriate energy average. Secondly it is proposed that the matrix elements of
the residual interaction are random, so that their average is zero. A formalism
incorporating these ideas, borrowed from statistical reaction theory, was developed
and explicit expressions for the mean field and the `fluctuation' away from the
mean field was obtained. An expansion of the fluctuation energy in terms of
increasing excitation complexity leads, after averaging, to formulas for the
corresponding contributions to fluctuation energy.

These formal discussion were followed in sect.~\ref{sec:HF} and 
\ref{sec:spectr-fact} and \ref{sec:numerical-results} by a simplified version and
by a schematic model. The most remarkable result is a sharp reduction the
effective strength of the residual interaction. But in addition the 
reasonableness of our results is very encouraging.

Of course much remains to be done. The quantitative connection with the 
underlying
nuclear forces has not been exhibited. The evaluation of the matrix elements
for finite  nuclei was not carried out and a better understanding of the 
energy average needs to be achieved. We have so far only a schematic model. What
is needed is a complete and thorough evaluation. But what has been indicated is
that such an evaluation will be successful.

Application to excited states is also indicated. In this case the smoothing
function used in reaction theory can be used instead of the one of section
\ref{sec:corrections}. This would lead to a complex mean field and the excited
state would have a width corresponding to the probability of the splitting of the
state by the residual interaction. The width would then measure the extent of the
splitting.

\appendix
\section {The bare interaction and the HF theory}
\label{app:app}

To implement the program outlined in Section \ref{sec:numerical-results}
we need a NN interaction to fix the HF energy of
nuclear matter.
 For illustrative purposes we choose the following simple NN interaction
\begin{equation}
 {\cal V}(r) = g_A \, \frac{e^{-\mu_A r}}{r}\, - 
   g_B\, \frac{e^{-\mu_B r}}{r} \, \frac{ 1 + P_x }{2} \,  ,
\label{nninter}
\end{equation}
which embodies a short-range repulsion in the first term and an intermediate 
range attraction in the second one (all the parameters are positive and
we require $\mu_A > \mu_B$).
The latter is taken of Majorana type, hence the occurrence of
the exchange operator
\begin{equation}
P_x = \frac{1 + \bbox{\tau}_1\cdot\bbox{\tau}_2}{2}\ 
\frac{1 + \bbox{\sigma}_1\cdot\bbox{\sigma}_2}{2}\,
\label{pxmajo}
\end{equation}
built out of the spin $\bbox{\sigma}$
and isospin $\bbox{\tau}$ operators. Thus (\ref{nninter})
contains the main features needed to account for the saturation
of the nuclear forces, reflected in  the existence of a minimum 
in the binding energy per particle $(B.E./A)$ versus $k_F$ curve,
but for the tensor force that here, for the sake of simplicity, is neglected.

The HF energy for the interaction  (\ref{nninter}) is easily worked out and 
leads to the following expression for the binding energy per particle 
\cite{Alb80}
\begin{eqnarray}
  \frac{B.E.}{A} & = & \frac{3}{5} \frac{\hbar^2 k_F^2}{2m} + 
    \frac{\varrho}{2}\, \left ( 4\pi\frac{g_A}{\mu_A^2} - 
    \frac{3\pi}{2} \frac{g_B}{\mu_B^2} \right ) 
     - \frac{3k_F}{4\pi} \left ( g_A + \frac{3}{2} g_B \right ) +
    \frac{1}{8\pi k_F} \left ( g_A \mu_A^2 + \frac{3}{2} g_B \mu_B^2 \right )
    \nonumber\\
  & & +\frac{1}{\pi} \left [  g_A \mu_A \, 
    \arctan \left (\frac{2k_F}{\mu_A}\right )
    + \frac{3}{2}   g_B \mu_B \,  \arctan \left (\frac{2k_F}{\mu_B}\right ) 
    \right ] \label{binden} \\
  & & - \frac{1}{8\pi k_F} \Biggl [  g_A \mu_A^2 
    \left( 3 + \frac{\mu_A^2}{4k_F^2}\right )
    \, \log \left ( 1 + \frac{4 k_F^2}{\mu_A^2}\right ) 
    + \frac{3}{2}   g_B \mu_B^2  \left( 3 + \frac{\mu_B^2}{4k_F^2}\right )
    \, \log \left ( 1 + \frac{4 k_F^2}{\mu_B^2}\right )\Biggr ] \, ,
\nonumber
\end{eqnarray}
where $\varrho = {2 k_F^3}/{3\pi^2}$.

The parameters characterizing the potential (\ref{nninter}) might be fixed,
for example, by accounting for the ``experimental'' nuclear matter values 
previously quoted.
One succeeds in doing so with the following choice
\begin{mathletters}
\begin{eqnarray}
\mu_A & = &  3.43 \ \text{fm}^{-1} \, ,\qquad \qquad  
\mu_B =   1.63 \ \text{fm}^{-1}
\label{muamub} \\
g_A&  = & 2460 \ \text{MeV~fm}\, ,\qquad \,  g_B  =  898 
\ \text{MeV~fm}\, .
\label{gagbexa}
\end{eqnarray}
\end{mathletters}

Worth noticing is that the range of the repulsion obtained with the 
fitting procedure is rather close to the one associated with the exchange of a 
$\omega$ meson  (3.97 fm$^{-1}$), whereas the range of the attraction
turns out to be intermediate to the one arising from the exchange of a pion
and of a $\sigma$ meson (0.71 fm$^{-1}$ and 2.79 fm$^{-1}$, respectively).

\begin{figure}[p]
\begin{center}
\mbox{\epsfig{file=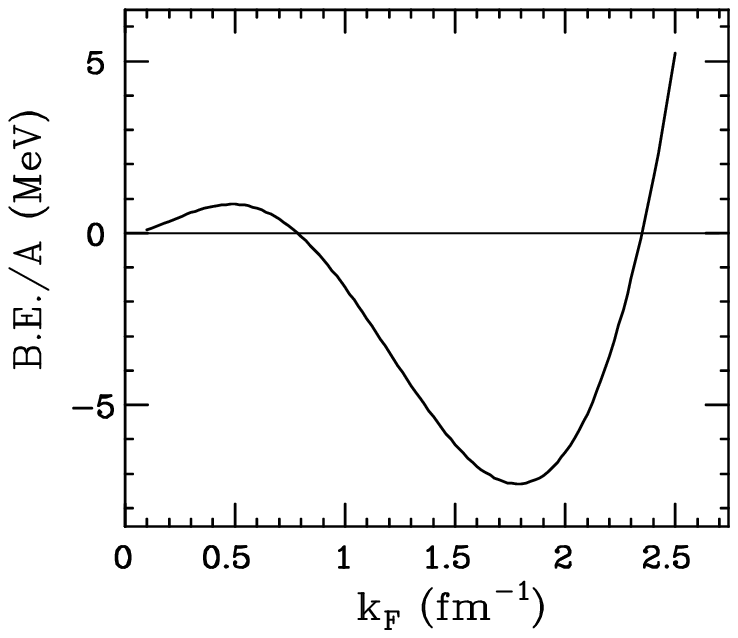}}
\vskip 3mm
\caption{
The binding energy per particle in the HF approximation (formula
(\protect\ref{binden}) of the text) for the potential (\protect\ref{nninter})
and the parameters given by (\protect\ref{gagbapp}).}
\label{fig:HF}
\vskip 5mm
\mbox{\epsfig{file=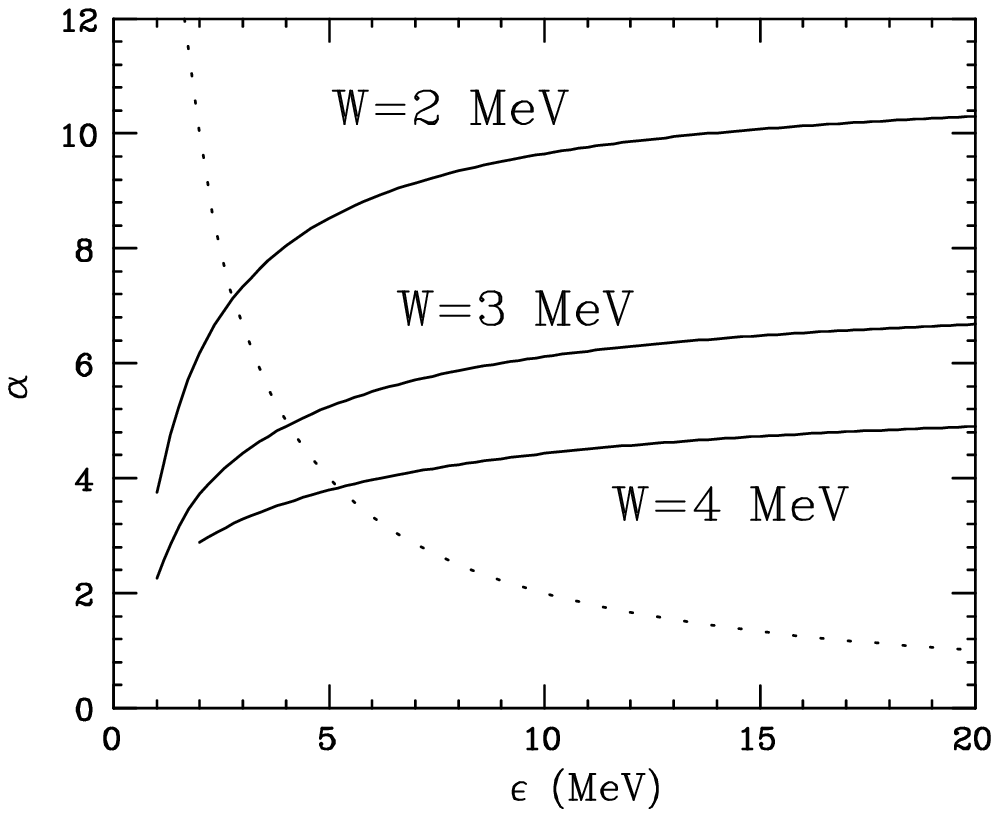}}
\vskip 3mm
\caption{ The loci corresponding to $\bar{E}_0^l=\bar{E}_0^u$ in the plane 
($\alpha,\epsilon$), for $W=2$, 3 and 4 MeV. Also shown (dotted line) 
is the curve along which the average energy of the 2p--2h excitations is  20 MeV.
}
\label{fig:alfeps}
\end{center}
\end{figure}

Here, we rather prefer to choose the parameters in such a way to have too 
little binding energy at too large a density in the HF frame, in order to 
conform to a shortcoming common to many nuclear matter calculations, the 
purpose being to ascertain whether the present theory is capable to improve 
upon the HF results.
Of course, there exists a variety of ways for reaching this scope:
In view of the rather realistic values of the ranges $\mu_A$ and $\mu_B$
(see (\ref{muamub})) we change the coupling constants.
We thus take, as a rather extreme  example,
\begin{equation}
g_A = 740\ \text{MeV~fm} \, ,\qquad g_B = 337 \ \text{MeV~fm}\, ,
\label{gagbapp}
\end{equation}
which, together with the values for $\mu_A$ and $\mu_B$ given in (\ref{muamub}),
yields a minimum of $-7.30$ MeV for the binding energy at $k_F=1.78$ fm$^{-1}$, 
as it can be seen in Fig.~\ref{fig:HF}, where the HF energy is displayed as a 
function of the density.

\section{Vacuum $\to$ 2p--2h matrix element}
\label{app:appb}

In the language of second quantization the matrix element we have to calculate
reads
\begin{equation}
 \beta^2 = \sum_{\scriptstyle\text{spin}\atop\scriptstyle\text{isospin}}
   \sum_{\scriptstyle k_1,k_2<k_F\atop
         \scriptstyle|\bbox{k}_1+\bbox{q}|,|\bbox{k}_2-\bbox{q}|>k_F}
  \langle\bbox{k}_1,\bbox{k}_2|V|\bbox{k}_1+\bbox{q},\bbox{k}_2-\bbox{q}\rangle
  \langle\bbox{k}_1+\bbox{q},\bbox{k}_2-\bbox{q}|V|\bbox{k}_1,\bbox{k}_2\rangle,
\label{eqb1}
\end{equation}
which, with standard manipulations, can be transformed into
\begin{eqnarray}
\beta^2 & = & A \frac{2}{\pi \varrho}
\int \frac{d \bbox{k}_1}{(2\pi)^3}\frac{d \bbox{k}_2}{(2\pi)^3} d \bbox{q}
\ \     \Theta \left ( |\bbox{k}_1 +\bbox{q} | -k_F \right )
\Theta \left ( |\bbox{k}_2 -\bbox{q} | -k_F \right )
\Theta \left ( k_F -k_1 \right ) \nonumber \\
&  &  \times\Theta \left ( k_F -k_2 \right ) 
\Biggl \{ 32 g_A^2 \frac{1}{(\mu_A^2 +q^2 )^2} +
12 g_B^2 \frac{1}{(\mu_B^2 +q^2 )^2}  
- 24 g_A g_B \frac{1}{\mu_A^2 +q^2} \, 
\frac{1}{\mu_B^2 +q^2} \nonumber \\
&  & - 8 g_A^2 \frac{1}{\mu_A^2 +q^2} \, \frac{1}{\mu_A^2 +| \bbox{k}_1 - 
\bbox{k}_2 + \bbox{q} |^2} 
+  12 g_B^2 \frac{1}{\mu_B^2 +q^2} \, \frac{1}{\mu_B^2 +| \bbox{k}_1 - 
\bbox{k}_2 + \bbox{q} |^2} \label{beta2} \\
&  &  -  12 g_A g_B \left ( \frac{1}{\mu_A^2 +q^2} \,  
\frac{1}{\mu_B^2 +| \bbox{k}_1 - \bbox{k}_2 + \bbox{q} |^2} + 
\frac{1}{\mu_B^2 +q^2} \, \frac{1}{\mu_A^2 +| \bbox{k}_1 -
\bbox{k}_2 + \bbox{q} |^2} \right ) \Biggr \} \, \nonumber 
\end{eqnarray}
(the matrix elements in eq.~(\ref{eqb1}) are understood to be antisymmetrized).
The above 9-dimensional integral,
by appropriate transformations, can be
reduced to a combination of 2- and 4-dimensional integrals, which can be
numerically evaluated, yielding the results quoted in  
Tables~\ref{tab:beta2_2}--\ref{tab:beta2_4}.
These values turn out to be helpful in performing a comparison with later 
findings.

The direct contributions, namely the first three terms in the right hand side of
(\ref{beta2}), are analytically evaluated.
Considering, e.~g., the third one, one gets 
\begin{eqnarray}
A \frac{2}{\pi \varrho} &&
\int \frac{d \bbox{k}_1}{(2\pi)^3}\frac{d \bbox{k}_2}{(2\pi)^3} d \bbox{q}
\ \     \Theta \left ( |\bbox{k}_1 +\bbox{q} | -k_F \right )
\Theta \left ( |\bbox{k}_2 -\bbox{q} | -k_F \right )
\Theta \left ( k_F -k_1 \right ) \Theta \left ( k_F -k_2 \right )
\nonumber \\
&&\qquad\times\left\{ \frac{1}{\mu_A^2 +q^2} \, 
\frac{1}{\mu_B^2 +q^2} \right\} \nonumber \\
&& = A \frac{\varrho}{4 k_F} \Biggl\{ \frac{1}{\tilde \mu_A^2 - \tilde \mu_B^2} 
\left (\tilde \mu_A \arctan \tilde \mu_A -\tilde \mu_B \arctan \tilde \mu_B
\right ) \\
&& + \frac{9}{5} + \frac{17}{12} (\tilde \mu_A^2 + \tilde \mu_B^2 )
+ \frac{1}{4} (\tilde \mu_A^4 + \tilde \mu_B^4 ) +
\frac{1}{4} \tilde \mu_A^2  \tilde \mu_B^2 \nonumber \\
&& +\frac{1}{4} \frac{1}{\tilde \mu_A^2 - \tilde \mu_B^2}  \left [
 \tilde \mu_B^3 (3 + \tilde \mu_B^2 )^2 \arctan 
\left ( \frac{1}{\tilde \mu_B} \right ) 
- \tilde \mu_A^3 (3 + \tilde \mu_A^2 )^2 \arctan
\left ( \frac{1}{\tilde \mu_A} \right )
\right ] \Biggr\} \nonumber \, ,
\end{eqnarray}
where $\tilde \mu_{A,B} = {\mu_{A,B}}/{2 k_F}$.

The other terms are obtained with obvious substitutions.

The exchange contribution is somewhat more involved. As an example we consider 
(again the other pieces are obtained through obvious interchanges)
\begin{eqnarray}
  \beta^2_{\text{exc}} &=& A \frac{2}{\pi \varrho} 
\int \frac{d \bbox{k}_1}{(2\pi)^3}\frac{d \bbox{k}_2}{(2\pi)^3} d \bbox{q}
\ \     \Theta \left ( |\bbox{k}_1 +\bbox{q} | -k_F \right )
\Theta \left ( |\bbox{k}_2 -\bbox{q} | -k_F \right )
\Theta \left ( k_F -k_1 \right ) \Theta \left ( k_F -k_2 \right )
\nonumber \\
&&\qquad\times\left\{ \frac{1}{\mu_A^2 +q^2} \,
\frac{1}{\mu_B^2 + | \bbox{k}_1 - \bbox{k}_2 + \bbox{q} |^2}
 \right\} \nonumber \\
&=& \frac{A}{(2 \pi^2 )} \frac{2}{\pi^2 \varrho} \sum_{l=0}^\infty 
(- 1)^l (2l +1) \int_0^\infty dq\, \frac{q^2 }{\mu_A^2 +q^2}
\int_0^\infty dr\, r e^{-\mu_B r} j_0 (qr) \\
&& \qquad \times \left [ \int_0^{k_F} dk\, k^2 j_l (kr) 
\int_{-1}^1 dx\, P_l (x)\,  \Theta (q^2 + k^2 + 2 q k x - k_F ) 
\right ]^2 \nonumber \, , 
\end{eqnarray}
where a partial wave expansion, in terms of the Legendre polynomials $P_\l$ and
the spherical Bessel functions $j_\l$, has been performed. 
The above integral, for $q\ge 2k_F$ is easily calculated yielding
\begin{eqnarray}
  \beta^2_{\text{exc}} &=& A \frac{2}{\pi^4 \varrho}
\int_0^\infty dr\, \frac{e^{-\mu_B r}}{r^6}\, 
\Bigl ( \sin (k_F r) - k_F r \cos (k_F r) \Bigr )^2
\int_{2 k_F}^\infty dq\, \frac{q }{\mu_A^2 +q^2}
\sin (qr) \, .
\end{eqnarray}
The remaining integrals are numerically evaluated.

\begin{table}
\caption{ In the second row the value of $\bar{E}_0$ as a function of $k_F$ is
reported; in the third row one finds the bare value of $\beta^2$, as given by
eq.~(\protect\ref{beta2}), while in the fourth and fifth rows the
estimated renormalized values on the lower and upper bounds, respectively;
in the last two rows, the corresponding values of the spectroscopic factor are
displayed. The values in this table correspond to $W=2$ MeV, $\epsilon=3$ MeV
and $\bar{\epsilon}_2=22$ MeV.
 }
\label{tab:beta2_2}
\begin{tabular}{lccc}
      $k_F$ (fm$^{-1}$) & 1.2  & 1.36 & 1.5 \\
  \tableline
  $\bar{E}_0$ (MeV)   & -15.3$\div$-15.5 & -16.3  & -15.6$\div$-15.2 \\
  \tableline
  $\beta^2$   & 4.5$\cdot$10$^4$  & 5.2$\cdot$10$^4$  & 7.0$\cdot$10$^4$  \\
  $\beta^2_l$ (MeV$^2$/nucleon) & 27.0  & 26.3  & 22.8  \\
  $\beta^2_u$                   & 54.2  & 49.0  & 36.4  \\
  \tableline
  $S_l$       &  0.32  &  0.32  &  0.30  \\
  $S_u$       &  0.64  &  0.62  &  0.58  \\
\end{tabular}
\vskip 3mm
\caption{ As in table~\protect\ref{tab:beta2_2}:
The values in this table correspond to $W=3$ MeV, $\epsilon=4$ MeV
and $\bar{\epsilon}_2=20$ MeV.
 }
\label{tab:beta2_3}
\begin{tabular}{lccc}
      $k_F$ (fm$^{-1}$) & 1.2  & 1.36 & 1.5 \\
  \tableline
  $\bar{E}_0$ (MeV)   & -15.5$\div$-15.9 & -16.6  & -15.9$\div$-15.2 \\
  \tableline
  $\beta^2$   & 4.5$\cdot$10$^4$  & 5.2$\cdot$10$^4$  & 7.0$\cdot$10$^4$  \\
  $\beta^2_l$ (MeV$^2$/nucleon) & 36.1  & 35.3  & 30.7  \\
  $\beta^2_u$                   & 79.6  & 69.9  & 49.8  \\
  \tableline
  $S_l$       &  0.34  &  0.33  &  0.32  \\
  $S_u$       &  0.69  &  0.67  &  0.63  \\
\end{tabular}
\vskip 3mm
\caption{ As in table~\protect\ref{tab:beta2_2}:
The values in this table correspond to $W=4$ MeV, $\epsilon=5$ MeV
and $\bar{\epsilon}_2=19$ MeV.
 }
\label{tab:beta2_4}
\begin{tabular}{lccc}
      $k_F$ (fm$^{-1}$) & 1.2  & 1.36 & 1.5 \\
  \tableline
  $\bar{E}_0$ (MeV)  & -15.7$\div$-16.4 & -16.8  & -16.1$\div$-15.2 \\
  \tableline
  $\beta^2$   & 4.5$\cdot$10$^4$  & 5.2$\cdot$10$^4$  & 7.0$\cdot$10$^4$  \\
  $\beta^2_l$ (MeV$^2$/nucleon) & 45.7  & 44.7  & 39.2  \\
  $\beta^2_u$                   & 108.6 & 91.9  & 62.7  \\
  \tableline
  $S_l$       &  0.35  &  0.34  &  0.33  \\
  $S_u$       &  0.73  &  0.72  &  0.66  \\
\end{tabular}
\end{table}

\begin{figure}[p]
\begin{center}
\mbox{\epsfig{file=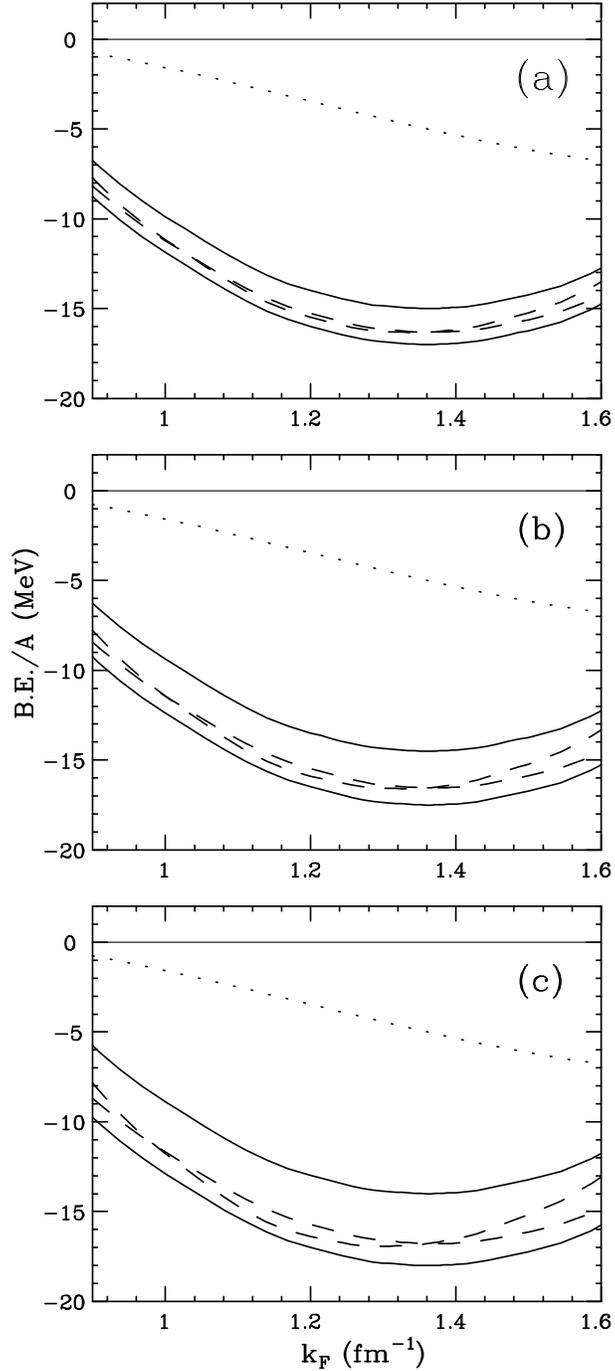,height=.8\textheight}}
\vskip 3mm
\caption{ The binding energy per particle in the HF approximation (dot) and in
the present approach: The solid lines represent the lower and upper bounds of
fluctuations for the energy $E$, whereas the dashed lines give the corresponding
mean field energies.
(a): $W=2$ MeV, $\epsilon=3$ MeV, $\alpha=7.3$, $\bar{\epsilon}_2 = 22$ MeV;
(b): $W=3$ MeV, $\epsilon=4$ MeV, $\alpha=4.9$, $\bar{\epsilon}_2 = 20$ MeV;
(c): $W=4$ MeV, $\epsilon=5$ MeV, $\alpha=3.8$, $\bar{\epsilon}_2 = 19$ MeV.
}
\label{fig:bind-energy}
\end{center}
\end{figure}

\end{document}